# Contrôle humain de l'Intelligence Artificielle et normalisation technique

Par Pr. Marion Ho-Dac & Dr. Baptiste Martinez
Univ. Artois, UR 2471, Centre Droit Éthique et Procédures (CDEP), F-59500 Douai, France

La prise de décision automatisée par des systèmes d'intelligence artificielle (« SIA ») peut nécessiter la mise en place de mesures de gouvernance particulières – distinctes de celles prévues pour des systèmes informatiques classiques – et ce, du fait du manque de transparence et du caractère non explicable de certains SIA[1]. L'adoption de mesures de contrôle humain en est une illustration majeure. De telles mesures permettent d'encadrer, à des degrés divers et suivant des modalités variées, le processus décisionnel des SIA, par exemple en plaçant une personne humaine en supervision du système et, en amont, en le développant techniquement pour permettre une telle supervision[2]. Par ricochet, les faiblesses des SIA concernés, y compris les conséquences négatives qu'ils pourraient avoir pour les personnes, devraient s'en trouver atténuées. Dans le même temps, plusieurs travaux ont montré que l'humain n'est pas toujours le garant d'un déploiement sans faille des SIA[3] ; si le contrôle humain a sans aucun doute pour vertu de renforcer la confiance des personnes affectées, il peut parfois s'avérer contreproductif face aux meilleures performances de la machine pour s'autocontrôler et atténuer les risques posés par le système[4]. À l'heure où l'Union européenne (« l'Union ») vient d'adopter l'un des premiers textes mondiaux visant à encadrer de manière horizontale et contraignante les SIA – l'AI Act –, on peut se demander comment le contrôle humain est réceptionné dans l'ordre juridique de l'Union[5].

---

[1] Sur le concept de décision dans l'écosystème numérique, v. L. Huttner, *La décision de l'algorithme. Étude de droit privé sur les relations entre l'humain et la machine*, Thèse, Paris 1-Panthéon-Sorbonne, 2022.

[2] Sur le contrôle humain, v. not. R. Crootof, M. E. Kaminski et W. N. Price II, « Humans in the Loop », *Vanderbilt Law Review*, 2023, p. 429 et s.; J. Laux, "Institutionalised distrust and human oversight of artificial intelligence: towards a democratic design of AI governance under the European Union AI Act", *AI & Soc*, 2023, https://doi.org/10.1007/s00146-023-01777-z ; W. Maxwell, *Le contrôle humain des systèmes algorithmiques – Un regard critique sur l'exigence d'un "humain dans la boucle"*, Mémoire HDR, Paris 1-Panthéon-Sorbonne, 2022 et « Le contrôle humain pour détecter les erreurs algorithmiques », in C. Castets-Renard & J. Eynard (dir.), *Un droit de l'Intelligence artificielle,* Bruxelles, Bruylant, 2023, p. 707 et s

[3] V. par ex. B. Green, "The flaws of policies requiring human oversight of government algorithms", *Computer Law & Security Review*, Volume 45, 2022, https://doi.org/10.1016/j.clsr.2022.105681

[4] En ce sens, J. Laux, "Institutionalised distrust and human oversight of artificial intelligence: towards a democratic design of AI governance under the European Union AI Act", *op. cit*. L'auteur relate que dans le domaine de l'aviation, par exemple, les opérations manuelles ont été de plus en plus limitées et remplacées par l'automatisation pour des raisons de sécurité.

[5] Règlement sur l'intelligence artificielle dit "AI Act" (non encore publié). Le texte utilisé est pour cette contribution est celui voté par le Parlement européen et ayant donné lieu à la résolution législative du 13 mars 2024 sur la proposition de règlement du Parlement européen et du Conseil établissant des règles harmonisées concernant l'intelligence artificielle (législation sur l'intelligence artificielle) et modifiant certains actes législatifs de l'Union (COM(2021)0206 – C9-0146/2021 – 2021/0106(COD)), P9_TA(2024)0138 (v. égal. version rectificative du 16 avril 2024).

Au sein de la gouvernance mondiale de l'IA[6], l'exigence de contrôle humain est concrétisée par plusieurs formats régulatoires, au sein d'une diversité de sources normatives : principes éthiques[7], normes techniques[8], règles de *soft law*[9] ou dispositions législatives contraignantes[10]. Dans sa dimension substantielle, le contenu donné au contrôle humain varie en fonction de « la place accordée à la personne humaine dans le processus décisionnel dans lequel le SIA intervient »[11]. Il s'agit d'un choix de politique juridique – celui des États et du législateur –, doublé d'un choix technique – à mettre en œuvre par les praticiens de l'IA, développeurs ou déployeurs des SIA notamment –. A l'aune du droit de l'Union antérieur à l'AI Act, les dispositions relatives au contrôle humain ont généralement vocation à régir des systèmes de prise de décision automatisée pouvant inclure un recours à l'IA, par exemple en matière de traitements de données personnelles[12] ou de modération des contenus en ligne[13]. Le contrôle humain intervient tantôt *ex ante* – avant la prise de décision – et donc durant la mise en œuvre du système, tantôt *ex post* – une fois la décision rendue par le système. L'exigence législative de contrôle humain permet, d'une part, de responsabiliser les utilisateurs desdits systèmes (par exemple, en leur imposant d'opérer certains contrôles) et, d'autre part, de protéger les personnes affectées par la décision (par exemple, en leur permettant de demander une révision de la décision).

Dans ce contexte, on peut affirmer que le droit de l'Union et, à travers lui, le législateur fixe la place de l'humain dans le processus décisionnel de l'IA. Toutefois, cette position législative est fragile car elle se limite à imposer un contrôle humain, sans en définir ni les contours, ni les outils. Que recouvre concrètement cette exigence, tant pour les utilisateurs, qu'en amont pour les fournisseurs de SIA et, en aval, pour les personnes affectées ? La loi ne le dit pas ; c'est aux opérateurs d'y pourvoir seuls ou au juge en cas de litige. De manière novatrice, l'AI Act – à partir duquel cette contribution raisonne –, ajoute une pierre notable à l'acquis législatif européen du contrôle humain. Il fait peser sur les fournisseurs de SIA à haut-risque (et dans une certaine mesure également sur les utilisateurs professionnels de ces systèmes qualifiés de déployeurs) des obligations dont la mise en place d'outils de contrôle humain, tout au long du cycle de vie des SIA, y compris *by design*[14] (et leur mise en œuvre par les déployeurs). Le législateur de l'Union va donc beaucoup plus loin que par le passé dans « l'explicitation » de l'exigence légale de contrôle humain. Mais le législateur n'entend pas pour autant tout prévoir ; il fait appel à la normalisation pour étoffer techniquement cette exigence (et plus largement l'ensemble des exigences de la section 2 du chapitre III) sur le fondement de l'article 40 de l'AI Act.

---

[6] Sur la notion de gouvernance régulatoire de l'IA, voy. M. Ho-Dac & C. Pellegrini, « Considering European consumer protection in the governance of artificial intelligence », *in* M. Ho-Dac & C. Pellegrini (ed.), *Governance of Artificial Intelligence in the European Union*, Brussels, Bruylant, 2023, p. 22. Comp. L. Floridi, « Soft Ethics and the Governance of the Digital », *Philos. Technol.*, 2018, vol. 31, p. 1-8.

[7] V. par ex. UNESCO, Recommandations sur l'éthique de l'IA, 23 novembre 2021 ; OCDE, Principes sur l'IA, 22 mai 2019. Plus largement, v. L. Floridi, (eds) *Ethics, Governance, and Policies in Artificial Intelligence*, Philosophical Studies Series, Springer, 2021, vol 144; J.-S. Gordon, S. Nyholm, "The Ethics of Artificial Intelligence, *in Internet Encyclopedia of Philosophy*, February 2021.

[8] Par ex. la norme en préparation ISO/IEC AWI 42105 Information technology- Artificial intelligence - Guidance for human oversight of AI systems. V. égal. les dispositions relatives au contrôle humain dans la norme IEEE *Standard for Transparency of Autonomous Systems* (IEEE Std 7001-2021), vol., no., pp.1-54, 4 March 2022.

[9] Par ex. les dispositions sur le contrôle humain prévues au sein du texte fédéral américain publié par The White House, *Blueprint for an AI Bill of Right: Making automated systems work for the american people*, 2022.

[10] Par ex. au Canada, la directive sur la prise de décision automatisée du 5 février 2019, spéc. § 6.3.11

[11] Il s'agit du point d'entrée de la recherche collective sur le contrôle humain proposée par la directrice du présent ouvrage, Nathalie Nevejans.

[12] V. article 22, §3 du règlement (UE) 2016/679 du Parlement européen et du Conseil du 27 avril 2016 relatif à la protection des personnes physiques à l'égard du traitement des données à caractère personnel et à la libre circulation de ces données, et abrogeant la directive 95/46/CE (règlement général sur la protection des données), JO L 119 du 4 mai 2016, p. 1-88.

[13] V. article 5, §3, *in fine* du règlement (UE) 2021/784 du Parlement européen et du Conseil du 29 avril 2021 relatif à la lutte contre la diffusion des contenus à caractère terroriste en ligne, JO L 172 du 17 mai 2021, p. 79-109.

Rappelons que la loi et les normes techniques fonctionnent généralement de manière complémentaire. Les normes (ou *standards* en anglais) sont des exigences volontaires, de nature technique ou de qualité auxquelles les produits, les processus de fabrication ou les services peuvent se conformer[15]. Elles sont rédigées par des organismes privés – tels que l'ISO[16], IEEE[17] ou, en Europe, CEN-CENELEC[18] et, en France, l'Afnor –, même si elles viennent souvent à l'appui d'une action législative ou en soutien d'une politique gouvernementale[19]. En matière d'IA, les normes techniques ont pour objectif le développement d'une IA de confiance, afin de guider les industriels dans l'intégration technique d'outils d'analyse, de prévention et d'atténuation des risques dont l'usage de l'IA peut être vecteur[20]. Plusieurs enceintes de normalisation travaillent actuellement dans le domaine de l'IA, sur le contrôle humain et d'autres notions connexes comme la transparence, l'explicabilité ou la question des biais algorithmiques[21]. Au niveau européen, la normalisation de l'IA est conduite par le *Joint Technical Committee 21* (« *JTC 21* ») au sein de CEN-CENELEC[22]. En vue de la demande de normalisation prévue par l'article 40 de l'AI Act, la Commission européenne a publié dès le mois de mai 2023 une demande de normalisation « anticipatoire » pour encadrer la préparation des futures normes harmonisées qui viendront compléter le règlement européen[23]. La future norme n° 5, en annexe de la demande de normalisation, est relative au contrôle humain[24].

---

[14] Cf. article 14 de l'AI Act, à lire en connexion avec d'autres dispositions satellites, telles que l'article 13 sur la transparence, l'article 9 sur la gestion des risques, l'article 11 sur la documentation technique (et l'annexe IV y relative) s'agissant des obligations du fournisseur et les articles 26 et 27 sous l'angle des obligations du déployeur.
[15] V. A. Penneau et D. Voinot, « Fasc. 970 : Normalisation », *J.-Cl. Concurrence-Consommation*, oct. 2010, mise à jour janv. 2023.
[16] Pour "*International Standardization Organization*" (organisation internationale de normalisation).
[17] Pour "*Institute of Electrical and Electronics Engineers*" (institut des ingénieurs électriciens et électroniciens).
[18] Le CEN (pour comité européen de normalisation) et le CENELEC (pour Comité européen de normalisation en électronique et en électrotechnique) sont des associations internationales sans but lucratif, officiellement reconnues comme organismes européens de normalisation, aux côtés de l'ETSI (pour Institut européen des normes de télécommunication). V. règlement (UE) n°1025/2012 du Parlement européen et du Conseil du 25 octobre 2012 relatif à la normalisation européenne, *JOUE* L 316, 14 novembre 2012, p. 12-33.
[19] See "Using and referencing ISO and IEC standards to support public policy", spec. p. 16. https://www.iso.org/files/live/sites/isoorg/files/store/en/PUB100358.pdf. V. égal. J. Dupendant, « International Regulatory Co-operation and International Organisations: The Case of the International Organization for Standardization (ISO) », OCDE/ISO, 2016 ; K. Jakobs, ICT Standardization, *Encyclopedia of Information Science and Technology*, 4th Edition, 2018, p. 13.
[20] V. CEN-CENELEC response to the EU White Paper in AI, June 2020 et CEN-CENELEC Focus Group Report: Road Map in Artificial Intelligence (AI) (en ligne).
[21] Pour une cartographie très complète et détaillée, v. European Commission, Joint Research Centre, Soler Garrido, J., Tolan, S., Hupont Torres, I. et al., *AI Watch – Artificial Intelligence Standardisation Landscape Update*, Publications Office of the European Union, 2023. Sous l'angle juridique et institutionnel, dans l'écosystème des TIC, v. O. Anevskaia, *The Law and Practice of Global ICT Standardization*, Cambridge University Press, 2023.
[22] V. P. Bezombes, « Key-role of standardisation in the field of artificial intelligence: how will future standards take consumer protection into account? », *in* M. Ho-Dac & C. Pellegrini (ed), *Governance of Artificial Intelligence in the European Union. What Place for Consumer Protection?*, Brussels, Bruylant, 2023, p. 171-185.

Dans ce contexte règlementaire multi-niveaux, la question de la place de l'humain dans le processus décisionnel de l'IA doit particulièrement retenir l'attention. En effet, selon que c'est la loi ou la norme technique qui fixe les contours du contrôle humain, la « gouvernance régulatoire » de l'IA n'a pas le même visage : sa nature, sa teneur et sa portée sont différentes. Trois principales caractéristiques distinguent, d'un côté, la loi et, de l'autre, les normes techniques (au-delà d'ailleurs du seul domaine de l'IA). La loi est adoptée par le biais d'un processus démocratique ; elle a une portée générale et contraignante (*hard law*) ; elle est gratuite et accessible à tous, généralement *via* sa publication officielle. La norme technique, par contraste, est adoptée par consensus lors d'un processus privé réunissant les parties prenantes ; elle a une portée technico-sociale et purement volontaire (*soft law*) ; elle est protégée par des droits de propriété intellectuelle et son accès est donc payant et ce – jusqu'à présent – même quand elle vient au soutien du droit de l'Union à travers le format particulier des normes harmonisées[25] dont seule la référence est publiée au Journal Officiel de l'Union Européenne[26]. Partant, on peut légitimement se demander comment répartir entre la loi et la norme technique la tâche de règlementer l'exigence de contrôle humain dans le contexte de l'IA.

Cette réflexion est au cœur de la contribution des juristes à la réflexion centrale sur la gouvernance régulatoire la plus appropriée – tout à la fois quant à son format institutionnel et à sa substance – pour assurer l'effectivité du contrôle humain au service d'une IA de confiance. Cela vaut particulièrement pour l'AI Act dont l'actualité, le large champ d'application et la probable influence mondiale en font un texte incontournable[27]. Comment l'AI Act a-t-il prévu de répartir l'exigence de contrôle humain entre ses dispositions de *hard law* et celles, volontaires, confiées à la normalisation ? Et cette répartition est-elle convaincante pour garantir un écosystème de confiance ?

---

[23] Pour une première analyse, v. J. Soler Garrido, D. Fano Yela, C. Panigutti, H. Junklewitz, R. Hamon, T. Evas, A. André and S. Scalzo, *Analysis of the preliminary AI standardisation work plan in support of the AI Act*, EUR 31518 EN, Publications Office of the European Union, Luxembourg, 2023. Dans la doctrine académique, v. notamment M. Ebers, "Standardizing AI – The Case of the European Commission's Proposal for an Artificial Intelligence Act", *in* L. DiMatteo, C. Poncibò, & M. Cannarsa, (Eds.), *The Cambridge Handbook of Artificial Intelligence: Global Perspectives on Law and Ethics,* 2022, Cambridge University Press; M. Ho-Dac, « La normalisation, clé de voute de la règlementation européenne de l'intelligence artificielle (*AI Act*) », *Dalloz IP/IT*, avril 2023, p. 228-233 ; Considering Fundamental Rights in the European Standardisation of Artificial Intelligence: Nonsense or Strategic Alliance? *in* K. Jakobs (Ed.), *Joint Proceedings EURAS & SIIT 2023*, Verlag Günter Mainz, 2023 ; A. Tartaro, "Towards European Standards Supporting the AI Act: Alignment Challenges on the Path to Trustworthy AI", *in* B. Müller (Ed.), P*roceedings of the AISB Convention 2023*, Swansea University 13/14 April 2023, p. 98-106.
[24] V. Commission implementing decision on a standardisation request to the European Committee for Standardisation and the European Committee for Electrotechnical Standardisation in support of Union policy on artificial intelligence, 22 mai 2023, Standardisation request M/593, C(2023)3215, cité ci-après « demande de normalisation en matière d'IA »
[25] Une « norme harmonisée » est définie comme « une norme européenne adoptée sur la base d'une demande formulée par la Commission pour l'application de la législation d'harmonisation de l'Union » par le règlement (UE) n°1025/2012 relatif à la normalisation, *op. cit.* Une telle norme est utilisée, par le droit dérivé aux fins de conférer une présomption de conformité aux produits devant être mis sur le marché aux exigences essentielles les concernant et établies par la législation d'harmonisation de l'Union. A cette fin, la référence des normes harmonisées est publiée au Journal Officiel de l'Union. Les opérateurs qui respectent les normes harmonisées reflétant les exigences essentielles prévues en droit dérivé suivant le mandat fixé par le texte bénéficient d'une présomption de conformité au droit dérivé.
[26] V. toutefois CJUE, 5 mars 2024, *Public.Resource.Org, Inc. et Right to Know CLG c/ Commission* (aff. C-588/21 P). Dans cet arrêt, la Cour impose (pour la première fois) la publication de la norme *in extenso* en matière de sécurité des jouets.

Deux axes analytiques, classiques en droit, sont proposés pour esquisser quelques éléments de réponse s'agissant de la place à accorder aux normes techniques, en contemplation de la loi européenne, en matière de contrôle humain. Il s'agit, en premier lieu, de l'étude de la notion de contrôle humain et, en second lieu, de celle de son régime.

**I) Les normes techniques à l'appui de la notion de contrôle humain de l'IA**

Afin d'apprécier la répartition opérée par l'AI Act entre loi et normalisation technique, la notion de contrôle humain doit être analysée doublement dans sa dimension intrinsèque et extrinsèque. Il s'agit, d'une part, d'explorer la notion de contrôle humain de l'intérieur en étudiant l'exigence de contrôle humain (A). D'autre part, en projetant la notion vers l'extérieur, les contours d'une « architecture notionnelle » du contrôle humain peuvent être identifiés (B).

**A) L'exigence du contrôle humain**

La notion de contrôle humain est à la fois technique, dans sa concrétisation au sein d'un SIA, et juridique, dans la mesure où la loi – à l'instar de l'AI Act – comme la jurisprudence y recourent[28]. Cette double dimension invite à déterminer si l'exigence de contrôle humain devrait plutôt être prévue par les normes techniques ou par les textes de loi. A cette fin, il faut établir ce qui, dans l'introduction d'une exigence juridique, relève de l'acte législatif, d'un côté, et de la norme technique, de l'autre. Dans l'ordre juridique de l'Union et par analogie, la Cour de justice de l'Union européenne (CJUE) s'est prononcée, plus largement, sur le domaine à réserver à la loi européenne, par rapport à d'autres actes normatifs non législatifs[29]. Elle a jugé que « l'adoption des règles essentielles de la matière envisagée est réservée à la compétence du législateur de l'Union »[30]. Ces règles essentielles « sont les dispositions dont l'adoption nécessite d'effectuer des choix politiques relevant des responsabilités propres du législateur de l'Union »[31] ou celles « qui ont pour objet de traduire les orientations fondamentales de la politique communautaire »[32]. La doctrine considère que c'est notamment le cas lorsqu'il s'agit

---

[27] V. not. M. Ho-Dac, « Législation européenne sur l'IA : point d'étape après la présidence française de l'Union européenne », *Dalloz IP/IT* 2022, p. 442 ; C. Lequesne, Adoption de l'*AI Act* : promesses et ambitions de la première législation occidentale sur l'intelligence artificielle, *Rec. Dalloz*, 2024, p. 864; J. Sénéchal, L'*AI Act* dans sa version finale – provisoire –, une hydre à trois têtes, *Dalloz Actu*, 11 mars 2024 ; M. Veale & F. Zuiderveen Borgesius, Demystifying the Draft EU Artificial Intelligence Act, *Computer Law Review International*, 4/2021, p. 97-112.

[28] Pour le droit de l'Union : CJUE, 21 juin 2022, *La Ligue des Droit humains* (aff. C 817/19) ; CJUE, 6 oct. 2020, *La Quadrature du Net* (aff. jtes C-511/18, C-512/18 et C-520/18). Pour le droit français : Cons. const., 12 juin 2018, 2018-765 DC ; Cnil, 22 mars 2018, 2018-119.

[29] Comp. pour une analyse critique de la normalisation comme délégation du pouvoir législatif, H-W. Micklitz, R. van Gestel, "European integration through standardization: How judicial review is breaking down the club house of private standardization bodies", (2013), 50, *Common Market Law Review*, Issue 1, pp. 145-181.

[30] CJUE, 5 sept. 2012, *Parlement/Conseil* (aff. C-355/10), pt. 64 ; v. égal. CJUE, 27 octobre 1992, *Allemagne/Commission* (aff. C-240/90), pt. 36

[31] CJUE, 5 sept. 2012, *op. cit.*, pt. 65.

[32] CJUE, 6 juillet 2000, *Molkereigenossenschaft Wiedergeltingen* (affaire C-356/97), pt. 21

de règlementer des questions relatives à la protection des droits fondamentaux[33]. Par comparaison, les normes sont définies en droit de l'Union comme des « spécifications techniques » à respecter par un produit, un processus, un service ou un système notamment quant à leurs niveaux de qualité, de performance, d'interopérabilité, de protection de l'environnement, de la santé ou de la sécurité[34]. Dans le même temps, on remarque que certaines normes récentes sont davantage tournées vers la protection des individus, de leurs droits fondamentaux et de la planète[35], comme le montrent les normes en matière de protection de la vie privée[36] et de devoir de diligence des entreprises[37]. Dans le domaine des technologies de l'information et de la communication, y compris de l'IA, on relèvera en outre que le rôle des normes est valorisé en réponse au principe de neutralité technologique que suit généralement le législateur et qui consiste à ne pas « viser trop précisément une technique particulière lors de l'élaboration de la loi […] [afin] de prévenir le risque d'obsolescence du droit »[38].

Transposé à l'exigence de contrôle humain, cet acquis permet de penser qu'il revient au législateur de l'Union, premièrement, de dire si un contrôle humain doit être prévu aux fins de mise sur le marché ou de mise en service d'un SIA à haut risque dans le for de l'Union, comme le fait l'AI Act [39]. Il s'agit d'un choix politique que de décider si un SIA peut ou non fonctionner seul et prendre des décisions automatisées opposables aux tiers dans tel ou tel domaine. En effet, l'exigence de contrôle humain est directement liée à la sauvegarde des droits fondamentaux des personnes affectées par le SIA dans la mesure où elle participe à réduire les conséquences négatives que la prise de décision automatisée pourrait avoir sur ces personnes[40]. Il s'agit, par exemple, de détecter des erreurs ou des discriminations dans le processus de décision, de modifier la décision finale en conséquence ou d'arrêter le système en cas d'anomalie grave. Quant à la norme technique, elle pourra intervenir en traduction et explicitation de la disposition législative introduisant le contrôle humain. La marge de manœuvre des opérateurs devrait en théorie rester entière pour aller, en application de la norme, au-delà de l'exigence légale de contrôle humain, par exemple en étendant son champ d'application à la mise sur le marché de SIA non couverts par l'AI Act[41].

Deuxièmement, c'est également, selon nous, à l'acte législatif de prévoir des éléments de définition, voire une définition générique, de l'exigence de contrôle humain qui pourrait être complétée techniquement dans une norme[42]. L'objectif est d'identifier clairement la notion, de façon à exclure des pratiques proches mais qui ne peuvent pas être assimilées à un contrôle humain, telles que la mise à l'arrêt d'un SIA sans considération de risque d'atteinte grave aux intérêts publics. Il devrait également s'agir d'opter clairement pour une conception donnée du contrôle humain face à la diversité des facettes attribuées à la notion. On peut ainsi regretter que cela ne soit pas le cas dans l'AI Act qui ne contient aucune définition expresse du contrôle humain[43]. Cela peut s'expliquer par le fait que la notion n'est pas suffisamment mature et prend des formes variables, selon ses contextes d'utilisation[44], rendant difficile tout essai de systématisation. De manière prospective, trois éléments principaux nous paraissent caractériser le contrôle humain suivant une approche englobante et ils pourraient servir de socle à sa définition : la compréhension humaine du SIA (au sens de maîtrise intellectuelle du système), la surveillance humaine du SIA (au sens de monitoring continu) et l'intervention humaine face au SIA (au sens de prise en main directe du système)[45]. L'AI Act intègre ces trois composantes mais elles figurent de manière éclatée dans le texte. Au regard de l'article 14 de l'AI Act et de son considérant 73, le contrôle humain pourrait être défini comme la capacité effective pour une personne physique qui a les compétences, la formation et l'autorité nécessaires, de superviser le fonctionnement du SIA tout au long de son cycle de vie, d'interpréter ses sorties et de lui donner des ordres, y compris de l'arrêter, au moyen d'interfaces homme-machine et de différents autres outils/procédés, y compris des mesures adoptées *by design*. Cette définition met en avant, de manière assez équilibrée, la triple dimension précitée du contrôle humain[46]. En écho au volet compréhension, l'article 26 de l'AI Act fixe des exigences de compétence et de maîtrise des SIA dans le chef des déployeurs. S'agissant du volet surveillance, l'accent est mis par l'AI Act sur la capacité du déployeur à interpréter les sorties du SIA[47] ; il s'agit d'une condition centrale en lien avec l'exigence de transparence afin de détecter des erreurs ou des biais et de pouvoir fournir des explications sur la prise de décision par le système aux personnes affectées. Ainsi, si l'on peut saluer la présence des éléments caractéristiques du contrôle humain dans l'AI Act, il est dommage que leur éparpillement au sein de plusieurs dispositions satellites ne soit pas compensé par une définition générique. Cela profiterait tant à l'intelligibilité de la règle de droit, qu'à la réception des dispositions législatives en matière de contrôle humain par les normes techniques.

---

[33] S. Thiery, *Les actes délégués en droit de l'Union européenne*, Bruxelles, Bruylant, 2020, p. 424.
[34] Article 2, §4 du règlement (UE) n°1025/2012 relatif à la normalisation, *op. cit.*
[35] Ce qui peut être source de tensions, comme le montre par ex. la consultation conduite en 2023 par le Haut-Commissariat aux Droits de l'Homme de l'ONU relative aux droits fondamentaux dans le processus de normalisation: https://www.ohchr.org/en/calls-for-input/2023/call-inputs-relationship-between-human-rights-and-technical-standard-setting
[36] V. EN 17529 *on data protection and privacy by design and by default* et ISO 27701:2019 – Techniques de sécurité. Extension d'ISO/IEC 27001 et ISO/IEC 27002 au management de la protection de la vie privée.
[37] ISO 26000 – Social responsibility.
[38] A. Latil, *Le droit du numérique. Une approche par les risques*, Paris, Dalloz, 2023, spéc. p. 30. Ce principe est qualifié de « neutralité juridique » par l'auteur et il distingue cette notion de la « neutralité des techniques » selon laquelle ces dernières seraient neutres d'un point de vue moral, social ou économique et, partant, ne pourraient pas être régulées par le droit, *op. cit.*, p. 22 et s.
[39] Article 14 de l'AI Act, *op. cit.*
[40] Considérant 66 et articles 8 et 9 de l'AI Act, *op. cit.*
[41] Il pourrait s'agir d'une stratégie normative « par le haut » pour les industriels qui commercialisent ou déploient des SIA à haut risque tout à la fois dans l'UE et hors de l'UE.

---

[42] Selon la CJUE, la définition ne doit pas nécessairement être complète mais doit être définie « avec suffisamment de précision » : CJUE, 14 octobre 1999, *Atlanta c/ Communauté européenne* (aff. C-104/97 P), pt. 76.

[43] Tel n'est pas non plus le cas dans les textes antérieurs précités consacrant le contrôle humain, ces derniers se limitant à prévoir un droit d'obtenir une intervention, une surveillance ou une vérification humaine.

[44] V. *supra*, dans l'introduction sur la diversité des « formats régulatoires » du contrôle humain et les références citées à ce propos.

[45] Comp. W. Maxwell, « Le contrôle humain pour détecter les erreurs algorithmiques », *op. cit.,* p. 711.

[46] Par comparaison, le rapport technique de l'ISO sur les préoccupations sociétales et éthiques en matière d'IA (ISO/IEC TR 24368:2022) propose des développements sur le contrôle humain de la technologie prévoyant qu'il a pour objectif de garantir que les décisions importantes restent soumises à un examen humain (*human review*). Il met donc l'accent sur le troisième volet notionnel du contrôle humain : l'intervention humaine. C'est à partir de cette dimension que les deux autres caractères – la compréhension et la surveillance – sont mobilisés. En ce sens, le rapport technique prévoit notamment que le contrôle humain implique, d'une part, d'avoir un humain qualifié dans la boucle, par exemple pour fournir l'autorisation de décisions automatisées et, d'autre part, d'évaluer de manière critique comment et quand déléguer des décisions aux systèmes d'IA, spéc. §6.2.8, ce qui renvoie aux premier et second volets notionnels que nous proposons.

[47] Articles 13, §3, sous d), article 14, Annexe IV, §3 sous e) et §3 de l'AI Act.

A côté de la loi, les normes ont également un rôle important à jouer en matière de terminologie[48] et, plus largement, de taxonomie dans le domaine de l'IA. La norme, en support de la loi, peut refléter ou compléter la définition légale de l'exigence de contrôle humain en la nourrissant techniquement. L'intérêt central d'une définition « normée », malgré son absence de force contraignante, est de dépasser les frontières législatives pour fournir une définition universelle du contrôle humain avec ses principaux éléments constitutifs, à la portée de tous les industriels, *designers*, développeurs et déployeurs dans le monde, peu importe leur ordre juridique d'appartenance. En ce sens, dans sa demande de normalisation, la Commission européenne indique que « la normalisation internationale peut contribuer à consolider une vision commune d'une IA de confiance à travers le monde, à faciliter le commerce et à éliminer les éventuels obstacles techniques »[49]. On peut donc souhaiter que la future norme harmonisée dédiée au contrôle humain contienne une définition de la notion comblant la lacune de la loi européenne. Cette définition sera néanmoins contrainte à l'aune des éléments de définition s'inférant de l'AI Act et devrait être pensée, dans le même temps, en dialogue avec l'orientation normative internationale[50].

**B) L'architecture notionnelle du contrôle humain**

Dans sa dimension extrinsèque, on constate que le contrôle humain n'est pas une notion « isolée ». Au contraire, elle apparaît comme englobante et en dialogue avec d'autres exigences régulatoires de l'IA de confiance. C'est ce que l'on propose de dénommer l'architecture notionnelle du contrôle humain. Cette architecture a une double dimension. La première, de type « macro », prend appui sur l'intégration du contrôle humain au sein de la gouvernance régulatoire de l'IA de confiance. La seconde, de type « micro », s'exprime par les interactions entre le contrôle humain et les autres exigences impératives applicables aux SIA dans l'AI Act. S'agissant d'abord de la « macroarchitecture notionnelle » du contrôle humain, elle est fondée sur le concept de l'IA de confiance. Ce concept a notamment été porté, dans l'Union, par le groupe d'experts de haut niveau sur l'IA (connu sous l'acronyme anglais « AI HLEG »)[51] et est largement embrassé par l'AI Act[52]. Selon le groupe d'experts, une IA digne de confiance est une IA licite, éthique et robuste. Pour parvenir à une IA de confiance, ce groupe propose aux praticiens de l'IA sept exigences essentielles – parmi lesquelles « l'action humaine et le contrôle humain » qui intègrent également le respect des droits fondamentaux[53] –, complétées par une liste d'évaluation de ces exigences. Dans ce contexte, le contrôle humain n'est donc pas seulement une exigence légale, applicable aux SIA à haut risque au sens de l'AI Act, mais il s'inscrit plus largement au sein de l'un des sept principes éthiques de l'IA de confiance, aux côtés de la robustesse et la sécurité ; le respect de la vie privée et la gouvernance des données ; la transparence ; la diversité ; la non-discrimination et l'équité ; le bien-être sociétal et environnemental ; et la responsabilité. Ces principes éthiques ont vocation à former un « cadre de haut-niveau »[54], en dépassement des exigences légales de la section 2 du chapitre III de l'AI Act. Le considérant 27 de l'AI Act invite les parties prenantes « à tenir compte […][de ces] principes éthiques pour l'élaboration de bonnes pratiques et de normes volontaires ». C'est la *soft law* (y compris les normes) qui devrait prendre ici le relai de la loi afin de dessiner une vision générique et transnationale du cadre régulatoire de l'IA de confiance[55]. En ce sens, certains experts au sein du *JCT 21* (préc.) travaillent à l'élaboration d'une norme-chapeau dessinant une « architecture normative de haut-niveau » des caractéristiques de l'IA de confiance[56]. Cette future norme devrait permettre d'expliciter chaque exigence impérative de l'AI Act, dont le contrôle humain, et d'organiser leurs interrelations, y compris au sein du cadre régulatoire général d'un SIA. La question reste ouverte de savoir comment positionner le contrôle humain en son sein : devrait-on en faire une caractéristique-pivot de ce cadre normatif de haut-niveau ? S'il ressort des travaux de l'AI HLEG que les principes d'éthique de l'IA ont une importance équivalente et ne sont pas hiérarchisés, la « micro-architecture notionnelle » du contrôle humain pourrait, à notre avis, apporter une réponse différente.

---

[48] Annexes à la demande de normalisation en matière d'IA, *op. cit.*
[49] V. demande de normalisation en matière d'IA, *op. cit.* et l'article 40, §3 de l'AI Act qui évoque « la coopération mondiale en faveur d'une normalisation en tenant compte des normes internationales existantes dans le domaine de l'IA […] ».
[50] Spéc. ISO/IEC AWI 42105 - Information technology — Artificial intelligence — Guidance for human oversight of AI systems (en cours de développement).
[51] Groupe d'experts indépendants de haut niveau sur l'intelligence artificielle (AI HLEG), *Lignes directrices en matière d'éthique pour une IA digne de confiance*, avril 2019.
[52] On compte plus d'un dizaine d'occurrences de l'expression « IA digne de confiance » dans l'AI Act.
[53] Lignes directrices, *op. cit.*, §58. Ces principes sont cités par l'AI Act dans son considérant 27.

La « microarchitecture » du contrôle humain, ensuite, permet de réfléchir à la façon dont le contrôle humain en tant qu'exigence légale (et non plus principe éthique), interagit et se positionne vis-à-vis des autres exigences impératives de l'AI Act, afin d'identifier comment la norme pourrait réceptionner cette microarchitecture. Selon le §2 de l'article 14 de l'AI Act, le contrôle humain vise « à prévenir ou à réduire au minimum les risques pour la santé, la sécurité ou les droits fondamentaux qui peuvent apparaître » lors de l'utilisation d'un SIA, « en particulier lorsque de tels risques persistent malgré l'application d'autres exigences énoncées dans la présente section ». Il s'agit d'une mise en interaction explicite, par la loi, du contrôle humain avec les autres exigences de la section 2 du chapitre III, consacrant implicitement, selon nous, le rôle-pivot du contrôle humain dans sa microarchitecture notionnelle. Par exemple, le contrôle humain pourrait permettre d'identifier des dysfonctionnements des fonctionnalités d'enregistrement prévues par l'article 12 de l'AI Act ou des manquements à l'obligation d'exactitude prévue par l'article 15. A cette fin, il est soutenu par les exigences de transparence de l'article 13 puisque ces dernières doivent notamment permettre à la personne en charge du contrôle humain d'interpréter « facilement » les sorties du SIA et, plus largement, de décrypter son processus décisionnel.

Face à ces interactions entre contrôle humain et autres exigences impératives, il peut y avoir des incertitudes d'articulation, voire des tensions venant freiner le cercle vertueux de l'architecture régulatoire de l'IA de confiance. Aussi, la répartition du rôle de la loi et de la norme est centrale sur ce point. D'une part, lorsque le législateur prévoit une exigence de contrôle humain, il paraît nécessaire de la relier clairement avec d'autres dispositifs régulatoires de l'IA de confiance, ainsi que de la positionner face à des notions connexes. Il en va de l'impératif d'intelligibilité de la loi. En pratique pourtant ce n'est guère le cas puisque l'AI Act n'établit de liens directs entre le contrôle humain comme « règle pivot » qu'avec les exigences de transparence et de documentation technique, et ces liens sont unilatéraux[57]. La transparence favorise le contrôle humain et le contrôle humain renforce la gestion des risques mais il n'y a pas de renforcement mutuel prévu entre ces exigences. La norme technique, d'autre part, pourrait aider les fournisseurs et les déployeurs à articuler les différentes exigences impératives et notamment celle du contrôle humain au sein de cette architecture notionnelle. La norme comblerait ainsi la lacune de l'AI Act sur ce point. La « norme-chapeau » en cours d'élaboration au sein du *JTC 21*, évoquée plus haut, pourrait y participer. De même, au sein de la future norme n°5 sur le contrôle humain, un schéma régulatoire d'interrelation entre les exigences connexes du contrôle humain devrait être élaboré et s'efforcer de préciser la teneur et les outils techniques de ces interrelations. Cela pourrait aussi passer par l'inclusion de profils verticaux sur le contrôle humain, c'est-à-dire suivant une approche sectorielle. La demande de normalisation de la Commission paraît aller en ce sens[58].

---

[54] V. ancien cons. 9 *bis* de la proposition d'AI Act amendée par le Parlement européen (non repris dans la version adoptée de l'AI Act) indiquant que les SIA « devraient tout mettre en œuvre pour respecter les principes généraux établissant un *cadre de haut niveau* qui favorise une approche cohérente et centrée sur l'humain d'une IA éthique et digne de confiance […]» en référence aux principes éthiques du AI HLEG. Comp. considérants 7 et 27 de l'AI Act.

[55] En ce sens le considérant 27 *in fine* de l'AI Act.

[56] « AI Trustworthiness framework » en cours d'élaboration.

L'analyse régulatoire du contrôle humain dans son volet notionnel a permis de montrer le rôle premier de la loi pour arrêter les contours notionnels de l'exigence de contrôle humain. Le relai de la norme est essentiel lorsqu'il s'agit d'étendre la portée notionnelle du contrôle humain au service de l'industrie mondiale de l'IA et d'agencer cette exigence dans l'architecture normative de l'IA de confiance. Qu'en est-il sur le terrain de la mise en œuvre du contrôle humain ?

**II) Les normes techniques à l'appui du régime du contrôle humain de l'IA**
Afin de poursuivre l'analyse de la répartition opérée par l'AI Act entre loi et normalisation technique, le régime du contrôle humain doit être analysé. Il permet, d'une part, d'établir que sa mise en œuvre est susceptible de degrés obligationnels (A) et, d'autre part, d'identifier ses outils techniques de concrétisation (B).

**A) Les degrés de mise en œuvre du contrôle humain**

Lorsqu'une exigence de contrôle humain est prévue, comme dans l'AI Act, la question se pose des degrés de ce contrôle lors de sa mise en œuvre. Il peut en effet être plus ou moins exigeant : on parlera de « degrés obligationnels » pesant sur le déployeur du SIA et imposant, en amont, au fournisseur du SIA de concevoir son système de manière appropriée[59]. L'identification de ces degrés obligationnels passe par une analyse systématique des textes législatifs (loi) ou

---

[57] On peut y ajouter, en dehors de la section 2 du chapitre III, l'article 27 en matière d'analyse d'impact sur les droits fondamentaux qui inclut la prise en compte des mesures relatives au contrôle humain (v. égal. cons. 91).
[58] Dans sa demande de normalisation précitée, la Commission européenne a précisé que les normes techniques européennes devraient prendre en compte les interdépendances entre les différentes exigences et, dans la mesure du possible, les refléter explicitement dans les spécifications correspondantes.
[59] Nous avons fait le choix de nous détacher de la typologie du AI HLEG sur le contrôle humain qui se place sur le terrain de la capacité d'intervention humaine, pour privilégier une approche régulatoire, c'est-à-dire fondée sur la prescription normative (de la loi et de la norme technique).

normatifs (norme) prévoyant un contrôle humain, analyse effectuée ici à l'aune de l'AI Act. A cette fin, la typologie suivante peut être utilisée :

Le degré peut être élevé, moyen ou faible, selon que la disposition sous analyse prévoit une exigence (« devoir »), une recommandation (« il convient de ») ou une possibilité (« pouvoir »)[60]. En outre, le degré obligationnel du contrôle humain prend un format de mise en œuvre tantôt *strict* au sens où la règle formulée est directement applicable, tantôt *variable* lorsque l'application concrète de la règle nécessite l'intervention du destinataire de l'obligation de contrôle humain. Dans ce dernier cas, ce n'est donc plus la règle qui fixe seule et de manière uniforme la mise en œuvre du contrôle humain. Cette typologie permet de porter une regard critique sur les dispositions règlementaires organisant le contrôle humain. On comprend que si son degré est élevé et le format de mise en œuvre strict, la teneur obligationnelle sera plus forte que si le degré est moyen et sa mise en œuvre variable.

Dans ce contexte, la question se pose à nouveau de la répartition régulatoire entre l'acte législatif et la norme technique. Fixer les degrés de contrôle humain relève d'abord de choix politiques consistant à réaliser un arbitrage entre l'efficacité de la technologie et la volonté de laisser une place à l'autonomie humaine. Il n'est donc pas uniquement question d'un processus rationnel reposant, par exemple, sur la nomenclature HABA-MABA (*Humans are better at / Machines are better at*)[61]. Il s'agit de poser des impératifs qui pourraient paraître « irrationnels » en termes d'efficacité des processus décisionnels mais qui sont démocratiquement définis au regard de valeurs sociétales et importants à l'aune des enjeux d'acceptabilité sociale de l'IA. S'agissant de la gouvernance européenne de l'IA, l'accent est mis, de manière explicite, sur une « IA axée sur l'humain »[62]. Cela implique de placer le bien-être et l'autonomisation de l'humain au centre de la technologie[63]. En prolongement, comme cela a été montré plus haut, le contrôle humain est rattaché à la protection des droits fondamentaux qui relève de la compétence de la loi[64]. Partant, il revient à l'acte législatif de définir les degrés obligationnels de contrôle humain. La norme technique a vocation à le compléter en précisant la mise en œuvre concrète du contrôle humain dans ses degrés obligationnels ; ces derniers pourraient être affinés, voire renforcés, par la norme mais non diminués.

Les articles 14 et 26 de l'AI Act fixent différents degrés obligationnels de l'exigence de contrôle humain. De manière générale, on remarque d'abord une domination du degré « élevé », ce qui est bienvenu eu égard à l'importance que revêt le contrôle humain en présence de SIA à haut risque. Les deux exceptions à la qualification de degrés « élevés » visent le déployeur. D'une part, il doit avoir la possibilité – sans y être directement obligé – de comprendre un certain nombre de fonctionnalités du SIA utilisé et leur mise en œuvre (degré moyen)[65]. D'autre part, il a « la faculté [...] d'organiser ses propres ressources et activités aux fins de la mise en œuvre des mesures de contrôle humain indiquées par le fournisseur » (degré faible)[66].

---

[60] Cf. typologie des expressions utilisées dans les Normes internationales ISO: https://www.iso.org/foreword-supplementary-information.html
[61] Sur cette nomenclature, v. not. R. Crootof, M. E. Kaminski et W. N. Price II, « Humans in the Loop », *op. cit.*
[62] V. considérants 1, 6, 8, 27, 176 et article 1er, §1 de l'AI Act.
[63] V. en ce sens la définition de l'IA centrée sur l'humain donnée par la « terminologie commune » US-EU TTC, oct 2023. https://digital-strategy.ec.europa.eu/en/library/eu-us-terminology-and-taxonomy-artificial-intelligence
[64] V. *supra*, I) A.
[65] Article 14, §4, sous a) à e) de l'AI Act.

Ensuite, quant au format de mise en œuvre, strict ou variable, des degrés obligationnels du contrôle humain, une domination du second format ressort de l'AI Act, ce qui pourrait affaiblir la teneur obligationnelle du contrôle humain. D'un côté, le législateur a posé une série de prescriptions strictes – de degrés obligationnels élevés – telles que la conception des SIA aux fins de permettre un contrôle effectif de leur déploiement par des personnes physiques[67] et l'exercice du contrôle humain par des personnes physiques compétentes, formées et ayant l'autorité et le soutien nécessaires[68]. En outre, on peut déduire de la teneur du contrôle humain que les déployeurs doivent être en capacité d'exercer, des prescriptions matérielles strictes à la charge du fournisseur, telles que : prévoir le suivi et le traitement des anomalies du SIA, prévenir les biais algorithmiques ou encore mettre en place un bouton d'arrêt[69]. En matière d'identification biométrique, le §5 de l'article 14 prévoit, par ailleurs, qu'aucune mesure ou décision automatisée ne doit être prise par le déployeur sans une double vérification humaine. Ces dispositions, en plus de leur degré obligationnel élevé, ont l'avantage grâce à leur format strict d'être matériellement explicites et directement actionnables, ce qui consolide le contrôle humain. De l'autre côté, l'AI Act prévoit une série de prescriptions qui vont varier sur la base d'un contrôle de proportionnalité exercé par le destinataire de la norme ou de la volonté de ce dernier. Cela s'explique par le fait que la définition même du contrôle est ancrée dans l'approche par les risques puisqu'il vise « à prévenir ou à réduire au minimum les risques » dans le déploiement des SIA[70]. La mise en œuvre du contrôle humain, par ricochet, implique des mesures « proportionnées aux risques, au niveau d'autonomie et au contexte d'utilisation » des SIA, à identifier par le fournisseur soit en les intégrant *by design*, soit pour être actionnées par le déployeur[71]. Le fournisseur peut donc moduler le régime juridique du contrôle humain sur la base, notamment, de son analyse des risques[72]. Une limite forte à ce pouvoir de modulation est prévue cependant, on l'a dit, en matière d'identification biométrique où une vérification humaine est imposée au titre des « mesures identifiées par le fournisseur » (degré strict)[73]. Enfin, une marge de manœuvre est reconnue au déployeur d'un point de vue organisationnel pour mettre en œuvre les mesures de contrôle humain prévues par le fournisseur[74].

---

[66] Article 26, §3 de l'AI Act.
[67] Article 14, §1 de l'AI Act.
[68] Article 26, §3 de l'AI Act. V. aussi l'utilisation par le déployeur du SIA en conformité avec la notice d'utilisation (y compris les exigences de contrôle humain) (26, §2) et l'article 4 de l'AI Act sur la maîtrise de l'IA.
[69] Artcile 16, §4 de l'AI Act.
[70] Article 14, §2 de l'AI Act.
[71] Article 14, §3 de l'AI Act.
[72] Art 14, §4 précisant « dans la mesure où cela est approprié et proportionné ». La version anglaise vise en outre les « circonstances ». Dans sa demande de normalisation, la Commission européenne indique pour sa part que les normes techniques européennes doivent prévoir, « lorsque cela se justifie, des mesures de contrôle appropriées spécifiques à certains SIA compte tenu de leur destination ».
[73] De manière critiquable, l'article 14, §5, alinéa 2 précise que cette règle stricte de double vérification humaine est exclue pour certains « sous » cas d'usage (justice pénale, migration et contrôle aux frontières) en matière de biométrie, lorsque le droit de l'Union ou le droit national la tiendrait pour disproportionnée. Cela fragilise le contrôle humain au détriment d'une approche pro-active des opérateurs de SIA à haut risque.
[74] Article 26, §3 de l'AI Act.

Au terme de cette typologie des degrés obligationnels du contrôle humain appliquée à l'AI Act, on peut se demander si le législateur de l'Union aurait dû aller plus loin dans l'édiction et l'explicitation des degrés obligationnels. On aurait effectivement pu souhaiter que la loi prévoie une liste plus systématique et précise des cas dans lesquels tel ou tel aspect du contrôle humain est prévu et selon quel degré obligationnel. De manière plus concrète, c'est le renforcement de ces degrés obligationnels dans les cas d'usage les plus sensibles à l'aune des droits fondamentaux des personnes concernées qui aurait été bienvenu[75]. Dans le même temps, on peut comprendre cette lacune qui s'explique certainement par la très grande diversité des destinations des SIA couverts par l'AI Act qui est pensé comme une règlementation horizontale et non sectorielle. Il est donc difficile pour la loi d'aller plus loin dans les détails de la mise en œuvre « obligationnelle » du contrôle humain au risque d'imposer des charges trop lourdes à l'ensemble de l'industrie de l'IA sans granularité suffisante. Il revient par conséquent à la norme technique de prendre le relai ; elle peut intervenir en complément de la loi, avec possiblement plus d'effectivité et de pertinence, pour poser des spécifications à la fois horizontales mais aussi verticales, propres à un secteur ou à un usage donné[76].

S'agissant de la réception par la norme des degrés obligationnels du contrôle humain tiré de l'AI Act, les dispositions législatives de degrés stricts ont vocation à être traduites et, le cas échéant, affinées techniquement au sein de la future norme harmonisée. Tel est par exemple le cas des exigences de compétence. La norme technique peut poser des critères de compétences professionnelles pour les personnes en charge du contrôle humain de SIA à haut risque. Il s'agit d'avancer dans la granularité du degré opérationnel strict en la matière s'agissant de la formation académique, de l'expérience professionnelle ou encore de critères d'aptitude qui seraient requis ou recommandés pour exercer les (ou certaines) prérogatives du contrôle humain en prolongement (notamment) l'article 26, §2 de l'AI Act qui vise « [les] compétence et [la] formation […] nécessaires »[77]. Par analogie, le *JTC 21* travaille actuellement sur un schéma de compétence des éthiciens de l'IA qui devraient être en charge de contrôler les caractéristiques de l'IA de confiance. Quant aux dispositions législatives de degrés variables, les normes techniques sont largement rompues à l'approche par les risques[78] associée à la définition même du contrôle humain. La question centrale réside donc dans l'appréciation du risque par la norme technique et la réponse à y apporter en termes de contrôle humain à l'aune des cadres généraux posés par la loi. A titre illustratif, la norme pourrait intégrer des méthodes permettant de mesurer et de répartir le juste niveau d'intervention humaine dans le processus de décision du SIA, à l'instar de la méthode de la performance de la décision[79]. Il s'agirait ainsi de répondre à la « variabilité » du degré obligationnel issue de la loi. Cette méthode consiste, selon ses concepteurs, à évaluer la pertinence du contrôle humain par rapport aux coûts (que sont les mauvaises décisions) et aux avantages (que sont les bonnes décisions) en fonction des probabilités que le SIA ou l'humain prennent ces bonnes ou mauvaises décisions. Plus le risque de mauvaises décisions par le SIA est élevé, plus un degré élevé d'intervention humaine est pertinent et, réciproquement, plus le risque de mauvaises décisions prises par l'humain est élevé, plus un degré d'autonomie décisionnelle du SIA est pertinent. Ces risques et avantages pourraient notamment être appréciés au regard de la protection des droits fondamentaux des personnes, dans la logique de la définition du risque au sens de l'AI Act.

---

[75] A titre illustratif, on aurait pu souhaiter qu'un degré élevé et de format strict du contrôle humain dans son volet « intervention humaine » (et plus particulièrement vérification de la décision par une personne) soit prévu par défaut pour l'ensemble des cas d'usage à haut risque en cas de risque d'atteinte à l'intégrité physique ou mentale de la personne.

[76] C'est en ce sens que se prononce la demande de normalisation de la Commission préc., §2.5, Annexe II.

[77] Des indications complémentaires sur le niveau de compétence / d'aptitude peuvent être déduites de l'article 14, §4 de l'AI Act aux fins de traduction/explicitation par la norme technique, par exemple lorsqu'il est prévu que le déployeur puisse interpréter, remplacer ou inverser les sorties du SIA, éviter les biais d'automatisation ou encore intervenir dans le fonctionnement du SIA. V. égal. article 4 de l'AI Act.

[78] Normes sur le management des risques (famille de normes ISO 31000) et AI Guidance on Risk Management, ISO/IEC 23894:2023.

**B) Les outils de mise en œuvre du contrôle humain**

Les normes techniques ont un rôle essentiel à jouer dans la détermination des outils de mise en œuvre du contrôle humain, dans la mesure où il s'agit de définir les procédés opérationnels par lesquels le contrôle humain est rendu praticable et effectif. On se situe donc principalement sur le terrain technique, en dehors des choix de politique juridique et d'arbitrage en matière de protection des droits fondamentaux propre au domaine de la loi ; ces choix ont assez largement été réalisés en amont, au stade des degrés obligationnels du contrôle humain. A ce titre, la norme devrait proposer des outils de mise en œuvre du contrôle humain qui couvrent les trois principaux volets du contrôle humain exprimés par l'AI Act (et précités plus haut), à savoir la compréhension humaine (au sens de maîtrise), la surveillance humaine (au sens de monitoring) et l'intervention humaine (au sens de prise en main). Quelques illustrations peuvent en être données.

S'agissant, premièrement, de la compréhension du fonctionnement du SIA, il s'agit fondamentalement pour la personne en charge du contrôle d'avoir les connaissances nécessaires lui permettant de superviser le système. A déjà été évoqué, à titre illustratif, l'idée d'un livrable normatif sur les compétences professionnelles des personnes physiques en charge du contrôle humain. Ce schéma de compétence peut également être lu à l'aune du système de management[80] du déployeur et il pourrait possiblement y être intégré. Le système de management de la qualité contient en principe un volet « compétence » et il est répliqué dans la norme ISO correspondante en matière d'IA[81]. Il prévoit notamment la nécessité de disposer de spécialistes ayant un bagage interdisciplinaire et experts dans les différentes étapes du cycle de vie du SIA. Si ces éléments sont intéressants, on peut souhaiter que la traduction européenne de cette norme en contemplation de l'AI Act soit plus précise quant aux exigences de compétence et ne vise pas seulement la performance de l'IA comme le fait la norme ISO précitée[82] mais également le contrôle humain. De manière plus spécifique et à la frontière entre la dimension de compréhension et celle de surveillance du SIA, l'article 14, §4, b) de l'AI Act prévoit que la mise en œuvre du contrôle humain doit permettre à la personne en charge d'avoir conscience du risque de biais d'automatisation. Ici le relai technique de la norme pourrait prendre la forme de spécifications informatives et explicatives relatives au contexte, conditions et expressions de tels biais[83], y compris une exigence de formation à ce type de biais[84]. Certaines de ces données pourraient ainsi être intégrées au schéma de compétence du contrôle humain, ainsi que dans la documentation technique du SIA.

---

[79] T. Baudel et G. Colombet et R. Hartmann, « ObjectivAIze : Déterminer empiriquement les rôles respectifs de l'humain et de l'algorithme dans la prise de décision », 33ème conférence internationale francophone sur l'Interaction Humain-Machine (IHM'22), AFIHM, avr. 2022, Namur, Belgique.
[80] Selon l'ISO, il s'agit de « l'ensemble des processus par lesquels un organisme gère les éléments corrélés ou en interaction de ses activités afin d'atteindre ses objectifs ». https://www.iso.org/fr/management-system-standards.html
[81] ISO/IEC 42001:2023 Technologies de l'information - Intelligence artificielle - Système de management
[82] Norme précitée, spéc. §7.2

Deuxièmement, la surveillance du fonctionnement du SIA consiste en la capacité de l'opérateur humain d'interpréter les sorties, de détecter et traiter les signes d'anomalies, de dysfonctionnements et de performances inattendues, voire même de pouvoir mener une enquête approfondie après un incident[85]. Les méthodes de surveillance que la norme technique pourrait réceptionner peuvent consister à réaliser un monitoring des performances en collectant en temps réel des données relatives au temps de réponse ou à l'utilisation des ressources. Cela peut aussi passer par l'intégration d'outils de gestion d'alertes permettant de signaler des anomalies concernant certaines métriques essentielles. A la frontière entre compréhension et surveillance du SIA, le contrôle humain dépend également des exigences en matière de transparence qui impliquent, selon l'AI Act, que les SIA à haut risque soient « conçus de manière à permettre aux déployeurs de comprendre [leur] fonctionnement, d'évaluer [leur] fonctionnalité et de comprendre [leurs] forces et [leurs] limites »[86]. Si la loi prévoit ces exigences[87], la norme a vocation à préciser techniquement les outils et méthodes pouvant permettre d'atteindre cette transparence souvent associée au concept d'explicabilité[88]. Le domaine de l'IA explicable a pour objectif de proposer des techniques permettant de comprendre et de superviser les systèmes d'IA. La norme IEEE P7001 en matière de transparence, par exemple, fixe une grille des mesures de transparence avec des niveaux croissants de complexité et de précisions, alliant approches documentaires[89], scénarios,

---

[83] La définition même du biais d'automatisation pourrait être proposée dans la norme, sur la base des éléments donnés dans l'AI Act (art. 14, §4 sous b) et en dialogue avec la défition retenue dans le document ISO/IEC TR 24027:2021 - Technologie de l'information - Intelligence artificielle (IA) - Biais dans les systèmes d'IA et dans la prise de décision assistée par IA, spéc. point 3.2.1. Contrairement à l'AI Act, ce texte prévoit que la tendance de l'humain à favoriser les résultats du SIA est complétée, en outre, par la propension « à ignorer les informations contradictoires produites sans automatisation, même si elles sont correctes ».

[84] Sous l'angle de la surveillance, il serait également possible de prévoir que l'interface du SIA rappelle à l'opérateur l'existence de ce risque de biais. Un message d'avertissement pourrait par exemple être affiché à l'écran.

[85] Suivant une lecture englobante du volet « surveillance » du contrôle humain, on pourrait ainsi y inclure l'obligation de plan de surveillance après commercialisation à réaliser par le fournisseur du SIA, notamment grâce aux données fournies par le déployeur (article 72) ainsi que les investigations que le fournisseur doit mener en cas d'incidents graves en dialogue avec le déployeur et les autorités de surveillance (article 73).

[86] Considérant 72 de l'AI Act.

[87] L'article 13 de l'AI Act prévoit que la conception et le développement des SIA à haut risque doivent être tels que leur fonctionnement est suffisamment transparent pour permettre aux utilisateurs d'interpréter les résultats du système et l'article 14, 4, c) indique que l'opérateur humain doit être en mesure d'interpréter correctement les résultats du SIA à haut risque, compte tenu notamment des caractéristiques du système et des outils et méthodes d'interprétation disponibles.

[88] C'est en ce sens que se prononce le rapport technique ISO 24368:2002 consacré aux enjeux éthiques et sociétaux en matière d'IA qui associe la transparence et l'explicabilité.

[89] Dans le domaine de la loi, le contenu de cette documentation est précisé par l'annexe IV de l'AI Act qui fait notamment mention d'une description détaillée des éléments du système d'IA et du processus de

principes de fonctionnement et matériels de formation interactifs[90]. Au plus haut niveau, elle propose également des techniques d'IA explicable. C'est également en ce sens que s'oriente le document en préparation ISO/IEC TS 6254 portant sur les objectifs et approches pour l'explicabilité des modèles d'apprentissage-machine et des systèmes d'IA. Reste que l'IA explicable est un domaine de recherche encore jeune et qui comprend certaines limites méthodologiques identifiées, telles que des incertitudes quant à la représentativité des explications fournies à l'aune du processus de décision du SIA[91]. L'IA explicable ne peut donc être qu'une des voies de transcription de la transparence du SIA aux fins de l'exercice du contrôle humain. Enfin, l'obligation de mener des investigations après un incident grave, telle que prévue par l'AI Act[92], soulève des interrogations en termes de protection des droits de propriété intellectuelle et de secret des affaires des fournisseurs de SIA. Partant, la norme technique devrait développer des schémas procéduraux permettant de mener des enquêtes tout en garantissant la protection de ces droits, par exemple sous l'angle de la confidentialité de certaines informations. Dans ce contexte, on précisera que la mise en balance des droits fondamentaux antagonistes des parties prenantes (fournisseurs, déployeurs et utilisateurs finaux) ne peut, en revanche, pas être traitée par la norme et relève du seul domaine législatif ou judiciaire.

Troisièmement, la mise en œuvre du contrôle humain devrait prendre la forme de mesures permettant une prise en main directe du SIA. Cette capacité d'intervention dans le processus de décision et, plus largement, dans le fonctionnement du SIA découle de l'AI Act et la norme devrait l'opérationnaliser. La loi précise déjà certaines modalités techniques que le fournisseur du SIA doit prévoir : permettre au déployeur de ne pas utiliser le SIA dans un cas donné ou de l'arrêter. La mise en œuvre de cette capacité d'intervention est du domaine de la norme technique qui pourrait notamment prévoir, d'une part, que l'interface permette un contrôle effectif et aisé, et d'autre part, que le SIA soit configuré de façon à ce qu'il ne puisse pas ignorer la commande humaine. S'agissant de l'interruption du SIA, l'AI Act vise explicitement le moyen d'un « bouton d'arrêt » ou, alternativement « une procédure similaire permettant au système de s'arrêter de manière sécurisée ». La norme technique doit préciser comment il est techniquement possible de le faire, par exemple en prévoyant un système de sécurité intégré apte à déclencher une interruption automatique en cas de détection d'anomalies graves.

L'analyse régulatoire du contrôle humain dans son volet mise en œuvre a permis de montrer le rôle premier de la loi dans l'établissement de « degrés obligationnels » dans les différentes mesures et procédés d'exercice du contrôle humain à la charge des fournisseurs et déployeurs de SIA. Ainsi, le contrôle humain joue à des degrés de rigueur variables, tout en laissant une marge de manœuvre opérationnelle aux opérateurs. La norme prend le relais pour approfondir techniquement la granularité de ces degrés obligationnels pour telle ou telle mesure de contrôle humain et pour les concrétiser à travers des outils techniques de mise en œuvre.

---

développement » ou encore d'« une description des éventuelles modifications apportées au système tout au long de son cycle de vie ».
[90] European Commission, Joint Research Centre, Soler Garrido, J., Tolan, S., Hupont Torres, I. et al., AI Watch – Artificial intelligence standardisation landscape update, *op. cit.,* p. 31.
[91] *Ibid*.
[92] Article 73, §6 de l'AI Act.

\* \* \* \*

En définitive, la complémentarité entre l'acte législatif et la norme technique est centrale dans le contexte de l'AI Act, afin d'atteindre un haut niveau d'effectivité du cadre régulatoire de l'IA dont l'exigence de contrôle humain constitue une pièce importante. Il en découle la nécessité d'une délimitation claire et d'une compréhension fine des domaines respectifs de la loi et de la norme. Un travail collaboratif doit être conduit au sein des organisations de normalisation de l'IA, entre, d'un côté les autorités publiques et, de l'autre, l'ensemble des parties prenantes, industriels et praticiens de l'IA, représentants de la société civile, chercheurs et universitaires. Cette collaboration ne doit pas s'arrêter aux frontières européennes mais embrasser la gouvernance régulatoire mondiale de l'IA, dans le respect des valeurs de l'Union. C'est ainsi que d'une notion et d'un régime européens du contrôle humain des SIA à haut risque pourra éclore une pratique transnationale du contrôle humain de l'IA de confiance.